\documentclass[prl, twocolumn, reprint, superscriptaddress, eprintnumbers]{revtex4-1}
\usepackage[]{graphicx}
\usepackage{epstopdf}
\usepackage{bm}
\usepackage{hyperref}
\usepackage{framed}
\usepackage{graphicx}
\usepackage{amsfonts}
\usepackage{wrapfig}
\usepackage{geometry}
\usepackage{sidecap,tikz, multirow,subcaption} 
\usepackage{braket, amsmath, amssymb, graphicx, tabu, lineno, tikz, xcolor, subcaption}

\begin{document}
	\title{Achieving Heisenberg limit in the phase measurement through three-qubit graph states}
	\author{Subhasish Bag **}
	\affiliation{Indian Institute of Technology Delhi, Hauz Khas, New Delhi, India - 110016.}

	\author{Ramita Sarkar **}\email{ramitasarkar11@gmail.com}
	\affiliation{Indian Institute of Science Education and Research Kolkata, Mohanpur, Nadia 741246, West Bengal, India}

	\author{Prasanta K. Panigrahi}
	\affiliation{Indian Institute of Science Education and Research Kolkata, Mohanpur, Nadia 741246, West Bengal, India}
	\date{\today}
	\begin{abstract}
		We study the reciprocal of the mean quantum Fisher information (RMQFI),  $\chi^2$ for general three qubit states, having graph and hypergraph states as special cases, for identifying genuine multi party entanglement characterized by  $\chi^2 <1$. We demonstrate that the most symmetric graph state and the GHZ state have the lowest RMQFI values leading to the highest statistical speed showing that both these states attain the Heisenberg limit in phase sensitivity. Unlike the GHZ state, graph states have the same RMQFI values for  measurement through different parameters, a property shared by the hypergraph states.  Three qubit graph and hypergraph states can violate Bell's inequality as $F_Q > N$. Both the GHZ state and the most symmetric graph state have the highest concurrence equalling 3 and the maximum QFI values. 
	\end{abstract}
	\maketitle
	\affiliation [** {Authors contributed equally}	

	\section{Introduction}
Quantum entanglement plays a significant role in the domain of quantum information. It can be physically described as the connection between two or more than two sub-systems of a multiparty quantum state and finds immense application in teleportation \cite{muralidharan2008perfect}, dense coding \cite{PhysRevLett.127.093601}, quantum state sharing \cite{gordon2006generalized} etc. The measure of quantum entanglement for  multipartite cases is of keen interest as it needed to be understood beyond bipartite case \cite{mansour2021bipartite}. Several measures such as geometric measure of entanglement \cite{chen2014comparison}, concurrence \cite{bhaskara2017generalized}, n-way distributive entanglement \cite{Banerjee_2020} are well investigated. Entanglement verification procedures like entanglement witnesses, Bell inequalities have been studied extensively\cite{guhne2009entanglement}. Quantum Fisher Information (QFI) can be treated as an important measure of entanglement. It is widely used in quantum sensing \cite{degen2017quantum},  quantum channel estimation \cite{katariya2021geometric}, quantum phase transition \cite{wang2014quantum,ye2016scaling,macieszczak2016dynamical}. QFI is also established as the fundamental notion of quantum metrology \cite{toth2014quantum,PhysRevResearch.4.013075,PhysRevResearch.3.023101}. Computing QFI is rather non-trivial. Cerezo et al. have analyzed e a lower bound on the QFI i.e. sub-quantum Fisher information (sub-QFI) \cite{cerezo2021sub}. Engaging generalized concepts of  fidelity and truncated states, another efficiently computable quantity has been introduced; known as truncated quantum Fisher information (TQFI) \cite{sone2021generalized}. Although estimating QFI for mixed state is computationally challenging, a variational quantum algorithm called variational quantum Fisher information estimation (VQFIE) is developed to carry out this task. QFI has been proved to be four times the convex roof of the variance \cite{PhysRevA.87.032324}. In 2009, Pezze and Smerzi introduced a quantity for N qubit state \cite{pezze2009entanglement}
\begin{equation}
	\chi^2= \frac{N}{F_Q[\rho_{in}, J_n]}
\end{equation}  where $F_Q[\rho_{in},J_n]$ is the quantum Fisher information. It is also shown that for multipartite entanglement, $\chi^2<1$. This criterion is useful for sub shot noise interferometry. An arbitrary N particle mixed separable state as an input of a linear interferometer cannot overcome the Shot Noise Limit (SNL) or standard Quantum Limit (SQL), $\Delta\theta=1/\sqrt{N}$ \cite{PhysRevA.55.2598,PhysRevLett.96.010401}. But Hyllus et al. have shown that all pure entangled states are not useful for SSN interferometry \cite{PhysRevA.82.012337}. They have established the relationship, $F_Q>N$ to archive the SSN sensitivity using the Cramer-Rao theorem. Considering Quantum Cramer Rao (QCR) bound $\Delta{\theta_{QCR}}=\frac{1}{\sqrt{F_Q}}=\frac{\chi}{\sqrt{N}}$, Pezze developed the condition for phase squeezing i.e $\chi<1$ \cite{pezze2009entanglement}. Recently, Heisenberg scaling has been obtained for frequency estimation in the presence of continuous time dephasing noise \cite{chiribella2022heisenberg}. {To parameterize physical quantity using quantum resources, the most fundamental limit for measuring precision is Heisenberg Limit (HL): $\Delta\theta_{HL}=1/N$ \cite{giovannetti2004quantum}}. 
 Phase sensitivity of experimentally observed bright solitary trains in Bose-Einstein condensate has been demonstrated too \cite{ghosh2019sub}.\\
Highly entangled multipartite pure states are efficient enough to use to develop quantum protocols. In this regard, graph states, hypergraph states, cluster states can be beneficial. Graph states, clusters can be utilized as various regular 2D lattices \cite{briegel2009measurement}. The criteria for the class of 2-colorable graphs are precisely studied, which is particularly useful in the context of entanglement purification \cite{hein2004multiparty}. The quantum hypergraph state \cite{sarkar2021geometry} was ﬁrst introduced in \cite{rossi2013quantum}, are highly entangled multiparty states \cite{dutta2019permutation}. Hypergraph states are the generalization of graph states as well as finite dimensional states. They have been largely implemented in various sectors such as quantum error correction \cite{schlingemann2001quantum,wagner2018analysis}, quantum blockchain \cite{banerjee2020quantum}, neural networks etc. Also, nonclassical properties of hypergraph states are exciting too \cite{sarkar2021phase}. Graph states can be a good resource for quantum metrology \cite{shettell2020graph}. Hypergraph states violate local realism  with
an exponential strength increasing with the number of particles are also predicted to  be beneficial for quantum metrology \cite{gachechiladze2016extreme}.\\
The outline of the paper is as follows. The description of graph and hypergraph and the corresponding graph and hypergraph state are given in section 2. RMQFI for the N particle system is described in section 3.  Section 4 is dealt with the calculation of RMQFI for a general 3 qubit along with the special example of graph and hypergraph state. Section 5 describes RMQFI with relative phase differences. After that, we came up with the concluding remarks.
\section{ Graph state and hypergraph state}
A graph $G = (V(G), E(G))$ can be expressed as a combination of vertex set V(G) and edge set E(G). A Loop is an edge connecting vertex with itself. Simple graph is defined as the graph which contains no loops and multiple edges. The vertices (a,b) are said to be adjacent when they are end points of an edge. An isolated vertex is defined as not adjacent to any other vertex.\\
hypergraph is the generalised version of graph, where an edge can connect any number of vertices, while the length of graph-edge is limited to 2. Both graph state and hypergraph state can be constructed from mathematical graph and hypergraph. \\
Here we constrain our discussion to 3 qubit graph and hypergraph state. Example of graph and corresponding graph state is given. We also give another example of the construction of 3 qubit hypergraph state.
\begin{figure}
	\centering
	\begin{subfigure}{0.4\textwidth}
		\begin{tikzpicture}[scale = 1]
			\draw[fill] (0, 0) circle [radius = 2pt];
			\node[above left] at (0, 0) {$3$};
			\draw[fill] (2.5, 0) circle [radius = 2pt];
			\node[above right] at (2.5, 0) {$2$};
			\draw[fill] (2.5, 2.2) circle [radius = 2pt];
			\node[above right] at (2.5, 2.2) {$1$};
			\draw [blue] (2.5, 2.2)--(2.5, 0);
			\draw [red] (2.5, 2.2)--(0, 0);
			\draw [green] (2.5, 0)--(0, 0);
		\end{tikzpicture}
		\caption{A graph with 3 vertices and 3 edges  connecting 1,2; 2,3; 3,1}
		\label{Fig1}
	\end{subfigure}
	~
	\begin{subfigure}{0.4\textwidth}
		\centering
		\begin{tikzpicture}[scale = 1.5]
			
			\draw (4, 0) -- (6.5, 0);
			\draw (4, 0.5) -- (6.5, 0.5);
			\draw (4, 1) -- (6.5, 1);
			\node at (3, 0) {qubit $1$};
			\node at (3, 0.5) {qubit $2$};
			\node at (3, 1) {qubit $3$};
			\node at (3.6, 0)  {$\ket{+}$};
			\node at (3.6, 0.5)  {$\ket{+}$};
			\node at (3.6, 1)  {$\ket{+}$};
			
			\draw[fill][blue] (4.7, 0) circle [radius = 2pt];
			\draw[fill][blue] (4.7, 0.5) circle [radius = 2pt];
			\draw [blue] (4.7, 0) -- (4.7,0.5);
			\draw[fill][red] (5.8, 0) circle [radius = 2pt];
			\draw[fill][red] (5.8, 1) circle [radius = 2pt];
			\draw [red] (5.8, 0) -- (5.8,1);
			\draw[fill][green] (5.25, 0.5) circle [radius = 2pt];
			\draw[fill][green] (5.25, 1) circle [radius = 2pt];
			
			\draw [green] (5.25, 0.5) -- (5.25,1);
		\end{tikzpicture}
		\caption{Quantum circuit for generating the graph state corresponding to the graph given in figure~\ref{Fig1}}
		\label{Fig2}
	\end{subfigure}
\end{figure}
The graph state corresponding to figure~\ref{Fig1} will be, 
\begin{equation} \label{graph1}
	\ket{G}=CZ_{(1,2)}CZ_{(2,3)}CZ_{(3,1)}\ket{+}^{\otimes{3}}    \end{equation}
Now, the corresponding state to \ref{graph1} is
\begin{equation}
	\begin{split}
\ket{G}=\frac{1}{\sqrt{8}}[\ket{000}+\ket{001}+\ket{010}-\ket{011}+
\\ \ket{100}-\ket{101}-\ket{110}-\ket{111}]
	\end{split}
\end{equation}

\begin{figure}
	\centering
	\begin{subfigure}{0.4\textwidth}
		\begin{tikzpicture}[scale = 1]
			\draw[fill] (0, 0) circle [radius = 2pt];
			\node[above right] at (0, 0) {$3$};
			\draw[fill] (2, 0) circle [radius = 2pt];
			\node[above left] at (2, 0) {$2$};
			\draw[fill] (2, 2) circle [radius = 2pt];
			\node[below left] at (2, 2) {$1$};
			\draw [blue] (-0.45,-0.3)--(2.3,-0.3)..controls (2.67,-0.09)..(2.5,0.65)--(2.25,2.4)..controls (2.15,2.6)..(2.,2.5)--(-0.5,0)..controls (-0.56,-0.2)..(-0.45,-0.3);
			
		\end{tikzpicture}
		\caption{A hypergraph with 3 vertices and only one hyperedge connecting 1,2,3}
		\label{Fig3}
	\end{subfigure}
	~
	\begin{subfigure}{0.4\textwidth}
		\centering
		\begin{tikzpicture}[scale = 1.5]
			
			\draw (4, 0) -- (6.5, 0);
			\draw (4, 0.5) -- (6.5, 0.5);
			\draw (4, 1) -- (6.5, 1);
			\node at (3, 0) {qubit $1$};
			\node at (3, 0.5) {qubit $2$};
			\node at (3, 1) {qubit $3$};
			\node at (3.6, 0)  {$\ket{+}$};
			\node at (3.6, 0.5)  {$\ket{+}$};
			\node at (3.6, 1)  {$\ket{+}$};
			
			\draw[fill][blue] (5.25, 0) circle [radius = 2pt];
			\draw[fill][blue] (5.25, 0.5) circle [radius = 2pt];
			\draw[fill][blue] (5.25, 1) circle [radius = 2pt];
			
			\draw [blue] (5.25, 0) -- (5.25,1);
		\end{tikzpicture}
		\caption{Quantum circuit for generating the hypergraph state corresponding to the hypergraph   given in figure~\ref{Fig3}}
		\label{Fig4}
	\end{subfigure}
\end{figure}
We have the mathematical expression for the hypergraph in figure \ref{Fig3}- 
\begin{equation} \label{h1}
	\ket{H}=C^2Z_{(1,2,3)}\ket{+}^{\otimes{3}}    
\end{equation}
Now, the corresponding state to \ref{h1} is
\begin{equation}
	\begin{split}
	\ket{H}=\frac{1}{\sqrt{8}}[\ket{000}+\ket{001}+\ket{010}+\ket{011}+\\ \ket{100}+\ket{101}+\ket{110}-\ket{111}]
	\end{split}
\end{equation}

\section{RMQFI for N particle system}
If we have the total N number of atoms and N$_1$ out of N are in the ground state and N$_2$ in the excited state then the corresponding state will be: $\ket{\frac{N}{2},m=\frac{N_2-N_1}{2}}=\ket{\frac{N}{2},m}$. Ex- In the case of 3-qubit GHZ state ($\frac{1}{\sqrt{2}}[\ket{000}+\ket{111}]$), there will be a superposition of $\ket{\frac{3}{2},\frac{3}{2}}$ and $\ket{\frac{3}{2},-\frac{3}{2}}$.\\
Considering $\rho_{in}$ is an input state and $J_n$ is the collective operator on the N particles, Pezzé and Smerzi \cite{pezze2009entanglement} introduced the quantity for an N qubit state :
\begin{equation}
	\chi^2=\frac{N}{F_Q[\rho_{in},J_n]}
\end{equation}

where, $F_Q[\rho_{in},J_n]=4(\Delta{R^2})$ is the mean quantum Fisher information. It was also shown for multipartite entanglement $\chi^2<1$. The Hermitian operator R is solved from \{R,$\rho_{in}$\}=i[$J_n$,$\rho_{in}$] \cite{boixo2008operational} where, the collective operator is defined as: $J_n=J.\Vec{n}$, $\Vec{n}$: an arbitrary direction. Also, $(\Delta{J}_{n+})^2_{max}$ is the maximum variance of a spin component $J_{n\perp}=J.{n}_{\perp}$ in the perpendicular plane to mean spin direction, we have for the pure state \cite{pezze2009entanglement}: $(\Delta{R^2})=(\Delta{J_{n\perp}})^2_{max}$. Then one can simplify the Reciprocal of the Mean Quantum Fisher Information per particle (RMQFI) as \cite{li2012dynamics} :
\begin{equation}
	\chi^2=\frac{N}{4(\Delta{J_{n\perp}})^2_{max}}
\end{equation}
The collective operator for N particles can be written as \cite{liu2010quantum}: 
\begin{equation}
	J_{\alpha}= \sum_{i=1}^{N} j_{i,\alpha} 
\end{equation}
Where $\alpha  {\in} x, y, z$ and $j_{i,\alpha}$ is the spin operator for the ith particle. A new coordinate frame was constructed for defining the collective operator mentioned in 
with the polar angle $\theta$, the azimuthal angle $\phi$ and the length of mean spin R as
\begin{equation}
	\theta=arccos(\frac{{\langle J_z \rangle}}{Rsin\theta}),
\end{equation}
\begin{equation}
	\phi=
	\begin{cases}
		arccos(\frac{{\langle J_x \rangle}}{Rsin\theta}) & \text{if ${\langle J_y \rangle}$ $\geq$ 0, }\\
		2\pi-arccos(\frac{{\langle J_x \rangle}}{Rsin\theta}) & \text{if $<J_y>$ $\leq$ 0}
	\end{cases}
\end{equation}
\begin{equation}
	R=\sqrt{{\langle J_x \rangle}^2+{\langle J_y \rangle}^2+{\langle J_z \rangle}^2}
\end{equation}
\begin{equation}
	{r=\sqrt{{\langle J_x \rangle}^2+{\langle J_y \rangle}^2}}
\end{equation}
With this new coordinate frame we can calculate the maximal variance \cite{yi2012quantum}
\begin{equation}
	\begin{split}
(\Delta{J_{n\perp}})^2_{max}=\frac{1}{2}\langle J_{n1}^2+J_{n2}^2 \rangle+\\ \frac{1}{2}\sqrt{{\langle J_{n1}^2-J_{n2}^2 \rangle }^2+ {\langle [J_{n1},J_{n2}]_+ \rangle}^2}
	\end{split}
\end{equation}
Where, 
\begin{equation}
	J_{n1}=sin\phi{J_x}+cos\phi{J_y},
\end{equation}
\begin{equation}
	J_{n2}=-cos\theta{cos\phi}J_x-cos\phi{sin\phi}J_y+sin\theta{J_z},
\end{equation}
\begin{equation}
	[J_{n1},J_{n2}]_+=J_{n1}J_{n2}+J_{n2}J_{n1}    
\end{equation}
\section{RMQFI of 3 qubit entangled State}
3 qubit general state can be written in the form of spin as follows, 
\begin{equation} \label{gen}
	\begin{split}
	\ket{\psi}=\alpha{e^{i\mu}}(\ket{\frac{3}{2},\frac{3}{2}})+\beta{e^{i\nu}}(\ket{\frac{3}{2},-\frac{3}{2}})+\\ \gamma{e^{i\eta}}(\ket{\frac{3}{2},\frac{1}{2}})+\delta(\ket{\frac{3}{2},-\frac{1}{2}})  	
	\end{split}
\end{equation}

If we change the basis from bit to spin we can write the state of 3-qubit graph state from \ref{graph1}:
\begin{equation}
\begin{split}
	\ket{G}=-\frac{1}{\sqrt{8}}\ket{\frac{3}{2},\frac{3}{2}}+\frac{1}{\sqrt{8}}\ket{\frac{3}{2},-\frac{3}{2}}-\sqrt{\frac{3}{8}}\ket{\frac{3}{2},\frac{1}{2}}\\ +\sqrt{\frac{3}{8}}\ket{\frac{3}{2},-\frac{1}{2}}
\end{split} 
\end{equation}
In the same manner we can write the state of 3-qubit hypergraph state from \ref{h1}:
\begin{equation}
	\begin{split}
	\ket{H}=-\frac{1}{\sqrt{8}}\ket{\frac{3}{2},\frac{3}{2}}+\frac{1}{\sqrt{8}}\ket{\frac{3}{2},-\frac{3}{2}}+\sqrt{\frac{3}{8}}\ket{\frac{3}{2},\frac{1}{2}}+\\ \sqrt{\frac{3}{8}}\ket{\frac{3}{2},-\frac{1}{2}}
\end{split} 
\end{equation}
The expectation values of the operators are:
\begin{equation}
	\langle J_x \rangle=\sqrt{3}\alpha\gamma{cos(\eta-\mu)}+2\gamma\delta{cos\eta}+\sqrt{3}\beta\delta{cos\nu}
\end{equation}
\begin{equation}
	 \braket{J_y}=\sqrt{3}\alpha\gamma{sin(\eta-\mu)}-2\gamma\delta{sin\eta}+\sqrt{3}\beta\delta{sin\nu}
\end{equation}
\begin{equation}
 \braket{J_z}=\frac{1}{2}[3({\alpha}^2-{\beta}^2)+{\gamma}^2-{\delta}^2]
\end{equation}
The generalized form can be depicted as,
\begin{equation}{\label{22}}
	\begin{split}
		\braket{J_{n1}^2+J_{n2}^2}=&\\
		\left[\frac{7}{4}-({\alpha}^2+{\beta}^2)\right]\left[1+\frac{\{3({\alpha}^2-{\beta}^2)+({\gamma}^2-{\delta}^2)\}^2}{4R^2}\right]&\\
		+\frac{\braket{ J_x}^2+{\langle J_y \rangle}^2}{R^2}\{\frac{1}{4}+2({\alpha}^2+{\beta}^2)\}&\\
		+\frac{\sqrt{3}}{R^2}({\langle J_y \rangle}^2-{\langle J_x \rangle}^2)\left[\gamma{\beta} \cos(\nu-\eta)+\alpha{\delta}\cos\mu \right]&\\
		-\frac{2\sqrt{3}}{R^2}{\langle J_x \rangle}{\langle J_y \rangle}[\gamma{\beta}\sin(\nu-\eta)-\alpha{\delta}\sin\mu]&\\
		-\frac{2\sqrt{3}}{R^2}{\langle J_y \rangle}{\langle J_z \rangle}[\alpha{\gamma}sin(\eta-\mu)-\beta{\delta}\sin\nu]-&\\
		\frac{2\sqrt{3}}{R^2}{\langle J_z \rangle}{\langle J_x \rangle}[\alpha{\gamma}cos(\eta-\mu)-\beta{\delta}\cos\nu]
	\end{split}
\end{equation}
\begin{equation}{\label{23}}
	\begin{split}
		\braket{ J_{n1}^2-J_{n2}^2}=&\\ \frac{{\langle J_x \rangle}^2+{\langle J_y \rangle}^2}{R^2}\{\frac{3}{2}-3({\alpha}^2+{\beta}^2)\}&\\
		+\sqrt{3}[1+\frac{\{3({\alpha}^2-{\beta}^2)+({\gamma}^2-{\delta}^2)\}^2}{4R^2}] &\\
		\frac{{\langle J_y \rangle}^2-{\langle J_x \rangle}^2}{{\langle J_x \rangle}^2+{\langle J_y \rangle}^2}[\gamma{\beta}cos(\nu-\eta)+\alpha{\delta}\cos\mu] &\\
		-2\sqrt{3}[1+\frac{\{3({\alpha}^2-{\beta}^2)+({\gamma}^2-{\delta}^2)\}^2}{4R^2}] &\\
		\frac{\langle J_x \rangle\langle J_y \rangle}{{\langle J_x \rangle}^2+{\langle J_y \rangle}^2}[\gamma{\beta}sin(\nu-\eta)-\alpha{\delta}\sin\mu] &\\
		+\frac{2\sqrt{3}}{R^2}\langle J_y \rangle\langle J_z \rangle[\alpha{\gamma}\sin(\eta-\mu)-\beta{\delta}\sin\nu] &\\
		+\frac{2\sqrt{3}}{R^2}\langle J_z \rangle\langle J_x \rangle[\alpha{\gamma}\cos(\eta-\mu)-\beta{\delta}\cos\nu]
	\end{split}
\end{equation}
\begin{equation}{\label{24}}
	\begin{split}
		&\langle [J_{n1},J_{n2}]_+ \rangle= \\ &4\sqrt{3}\frac{\langle J_x \rangle\langle J_y \rangle\langle J_z \rangle}{Rr^2}[\gamma{\beta}\cos(\nu-\eta)+\alpha{\delta}\cos\mu]\\
		&+2\sqrt{3}\frac{\langle J_z \rangle({\langle J_y \rangle}^2-{\langle J_x \rangle}^2)}{Rr^2}[\gamma{\beta}\sin(\nu-\eta)\\
		&-\alpha{\delta}\sin\mu]-\frac{2\sqrt{3}}{R}{\langle J_y \rangle}[\alpha{\gamma}\cos(\eta-\mu)-\beta{\delta}\cos\nu]\\
		&+\frac{2\sqrt{3}}{R}{\langle J_x \rangle}[\alpha{\gamma}sin(\eta-\mu)-\beta{\delta}\sin\nu]
	\end{split}
\end{equation}
The  reciprocal of the mean quantum Fisher information per particle (RMQFI) is plotted with respect to the parameters of the general 3 qubit state; the graph state and hypergraph state mentioned in the example are shown in the plot too. \\
\begin{figure}[h]
	\includegraphics[scale=0.15]{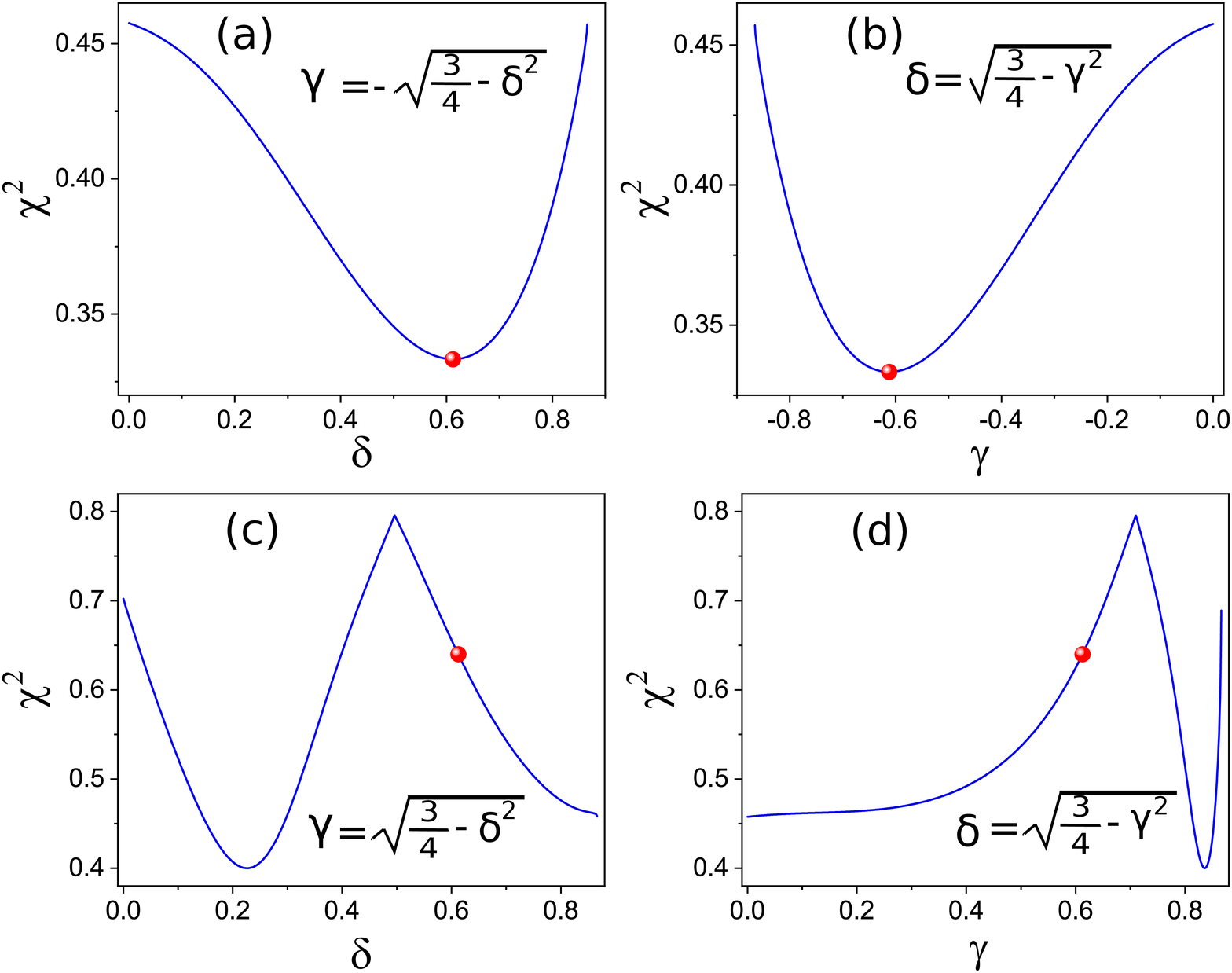}
	\caption{ $\alpha=-\frac{1}{\sqrt{8}}$, $\beta=\frac{1}{\sqrt{8}}$ and $\mu=\nu=\eta=0$, In figure a,b red point depicts graph state; while figure c,d red point indicates hypergraph state. In figure a, b $\gamma$ is negative while figure c,d $\gamma$ is positive. }
	\label{all}
\end{figure}\\
Now, the outcome of the variation  $\chi^2$ (RMQFI) with respect to the parameters of the equations \ref{gen} and results are mentioned in the table \ref{tab:label all table},\ref{tab:label graph table}.	

\begin{table}[h]
	\begin{subtable}[h]{0.35\textwidth}
		\centering
		\begin{tabular}{| p{2.2 cm} | p{2.5 cm} | }
			\hline
			Parameter $\delta$ & $\chi^2$ \\
			\hline 
			0.4960 & $\chi^2_{max}$=0.795775 \\
			\hline  
			0.2260 & $\chi^2_{min}$=0.399956 \\
			\hline  
			$\sqrt{\frac{3}{8}}$, hypergraph state in the example & 0.6400\\
			\hline
		\end{tabular}
		\caption{$\chi^2$ is varied with respect to $\delta$ and $\gamma=\sqrt{\frac{3}{4}-\delta^2}$}
		\label{tab:label subtable A}
	\end{subtable}
	\hfil
	\begin{subtable}[h]{0.35\textwidth}
		\centering
		\begin{tabular}{| p{2.2 cm} | p{2.5 cm} | }
			\hline
			Parameter $\gamma$ & $\chi^2$ \\
			\hline 
			0.7100 & $\chi^2_{max}$=0.795598 \\
			\hline  
			0.8360 & $\chi^2_{min}$=0.399955 \\
			\hline  
			$\sqrt{\frac{3}{8}}$, hypergraph state in the example & 0.6400\\
			\hline
		\end{tabular}
		\caption{variation of RMQFI with respect to $\gamma$ and $\delta=\sqrt{\frac{3}{4}-\gamma^2}$}
		\label{tab:label subtable B}
	\end{subtable}
	\hfill
	
	\caption{General region with parameters $\alpha:-\frac{1}{\sqrt{8}}, \beta:\frac{1}{\sqrt{8}}$, $\mu=\nu=\eta=0$ ; where $\chi^2$ is varied with respect to $\delta$ and $\gamma$ respectively. Maximum and minimum value of $\chi^2$ are given in the table, along with the value of $\chi^2$ of hypergraph state in example when varied with $\delta$ and $\gamma$ both}
	\label{tab:label all table}
\end{table}

\begin{table}
	\begin{subtable}[h]{0.35\textwidth}
		\centering
		\begin{tabular}{| p{2.2 cm} | p{2.5 cm} | }
			\hline
			Parameter $\delta$ & $\chi^2$ \\
			\hline 
			0.0 & $\chi^2_{max}$=0.45758 \\
			\hline  
			0.61237 & $\chi^2_{min}$=0.33333 \\
			\hline  
			$\sqrt{\frac{3}{8}}$, Graph state in the example & 0.3333\\
			\hline
		\end{tabular}
		\caption{$\chi^2$ is varied with respect to $\delta$ and 
			$\gamma=-\sqrt{\frac{3}{4}-\delta^2}$}
		\label{tab:label subtable C}
	\end{subtable}
	\hfil
	\begin{subtable}[h]{0.35\textwidth}
		\centering
		\begin{tabular}{| p{2.2 cm} | p{2.5 cm} | }
			\hline
			Parameter $\gamma$ & $\chi^2$ \\
			\hline 
			-0.0030 & $\chi^2_{max}$=0.45758 \\
			\hline  
			-0.61237 & $\chi^2_{min}$=0.33333 \\
			\hline  
			$-\sqrt{\frac{3}{8}}$, Graph state in the example & 0.3333\\
			\hline
		\end{tabular}
		\caption{$\chi^2$ is varied with respect to  $\gamma$ and 
			and $\delta=\sqrt{\frac{3}{4}-\gamma^2}$}
		\label{tab:label subtable D}
	\end{subtable}
	\hfill
	
	\caption{General region with parameters $\alpha:-\frac{1}{\sqrt{8}}, \beta:\frac{1}{\sqrt{8}}$, $\mu=\nu=\eta=0$ ; where $\chi^2$ is varied with respect to $\delta$ and $\gamma$ respectively. Maximum and minimum value of $\chi^2$ are given in the table, along with the value of $\chi^2$ of graph state in example when varied with $\delta$ and $\gamma$ both}
	\label{tab:label graph table}
\end{table}
\subsection{RMQFI for 3 qubit graph and hypergraph states}
We have studied QFI for general 3 qubit pure states in spin based system. We have observed different values of RMQFI, varying different parameters of the general states. keeping $\alpha:-\frac{1}{\sqrt{8}}$ and $\beta:\frac{1}{\sqrt{8}}$ fixed at these two values, RMQFI is measured for different regions varying the  coefficients $\gamma$ and $\delta$. By changing those parameters, general region of the RMQFI  for the two different cases: (I) $-\gamma$ and (II) $+\gamma$ are observed, keeping other parameters constants at certain values. From these two sets, we get the RMQFI$<$1, which implies the entangled region and pointing out the exact values for the graph and hypergraph states. \\
From the tables, we identified that both graph and hypergraph states have RMQFI values less than 1 implying multiparty entanglement. Also, it is found that the graph state always has the lowest RMQFI values i.e. 0.33, while the hypergraph state lies between maximum and minimum. Graph states being equally probable with respect to the parameters have the same RMQFI values while measured through any of the parameters. Also, hypergraph states behave similarly.\\
A relation between the QFI and statistical speed can be defined ($v_F$) as $v_F^2=F_Q$ \cite{pezze2009entanglement}. Statistical speed for the graph state $\ket{G}$ and hypergraph state $\ket{H}$ are $3$ and $2.165$ respectively. Statistical speed can be described as the relative distance between two states \cite{braunstein1994statistical,wootters1981statistical}. Considering the entangled criteria, the critical velocity $|v_{cr}|=\sqrt{N}$ will be overcome only by the entangled states, and the maximum speed $(v_{max}=N)$ corresponding to the strongest entangled state implies the Heisenberg Limit(HL) \cite{giovannetti2004quantum} in phase sensitivity: $\Delta\theta_{HL}=1/N$. Here, in this case, we found that this graph state has achieved the maximum speed, and the corresponding $\Delta\theta_{HL}$ is $\frac{1}{{3}}$.\\
\subsection{QFI and concurrence}	
Concurrence, being an important measure entanglement \cite{wootters1998entanglement} is strictly positive for
entangled states and vanishing for all separable states. This measure contains the entire entanglement, providing necessary and sufficient conditions for separability.
Concurrence for a bipartition $M|M^\prime$ is defined as \cite{sarkar2021geometry, dutta2019permutation, roy2021geometric},
\begin{equation}
	E^2_{M| M ^\prime}=2[1-tr{(\rho^M)^2}]
\end{equation}
Where $\rho^M$ is the reduced density matrix on the sub-system M.
3 qubit GHZ having the concurrence value 3 reached the Heisenberg limit in phase sensitivity $\Delta\theta_{HL}=1/N=1/3$. The most symmetric graph state given in fig-\ref{Fig1}, also has the concurrence value 3 and $\Delta\theta_{HL}$ as $1/3$. GHZ state, as well as this graph state, have the same and highest QFI value i.e. 9. While this 3 qubit  hypergraph state having lower concurrence 2.5981, has lower QFI 4.6875. {So, it is easy to observe higher concurrence value implies higher QFI value.}
\section{RMQFI with relative Phase Differences}
\begin{figure*}[h]
	\centering
	\includegraphics[scale=0.95]{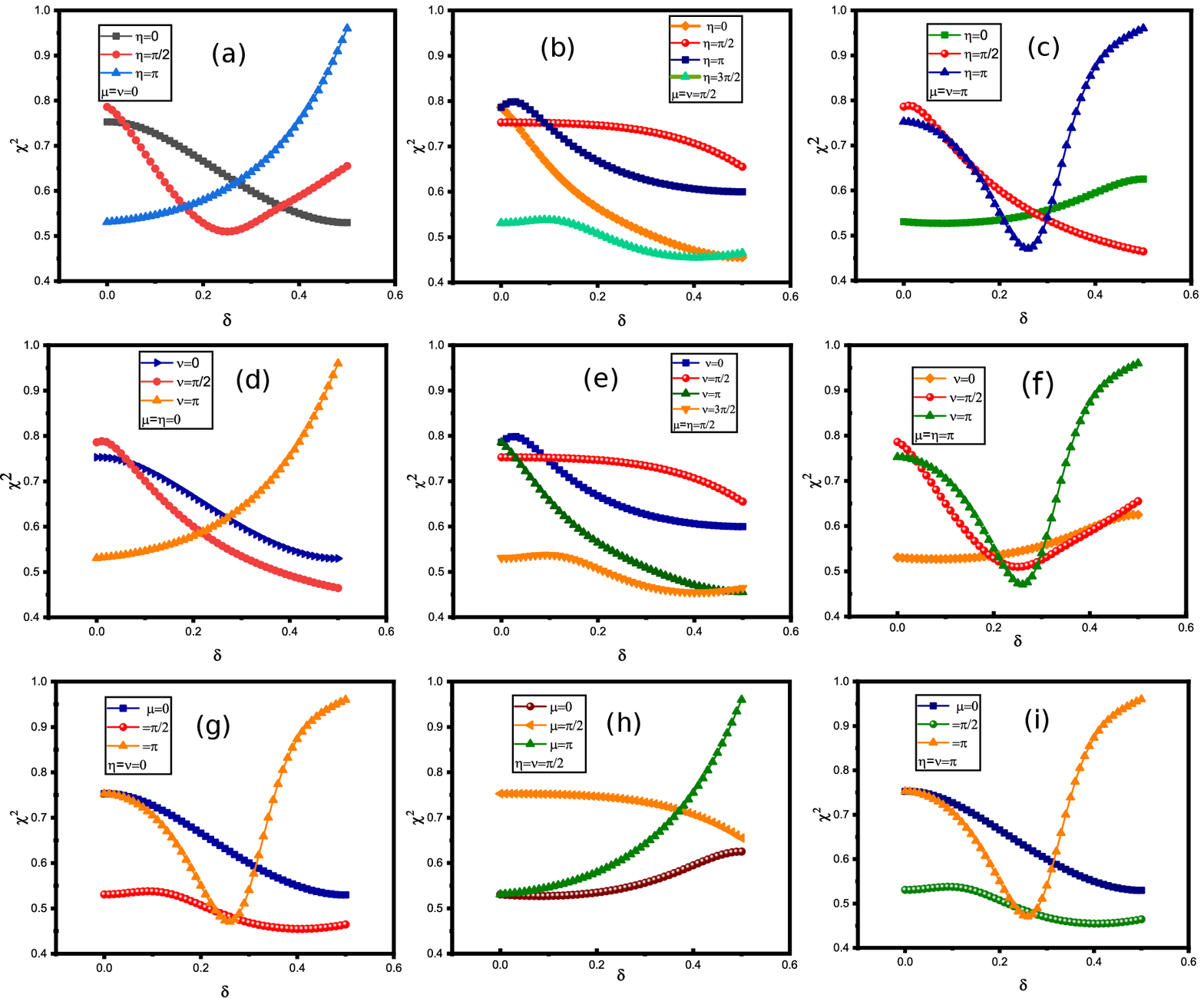}
	\caption{ Variation of $\chi^2$ with respect to $\delta$, keeping $\alpha=\beta=\frac{1}{2}$ and for different phase values of $\mu,\nu,\eta$.}
	\label{phase_1}
\end{figure*}

\begin{figure*} 
	\centering
	\includegraphics[scale=0.95]{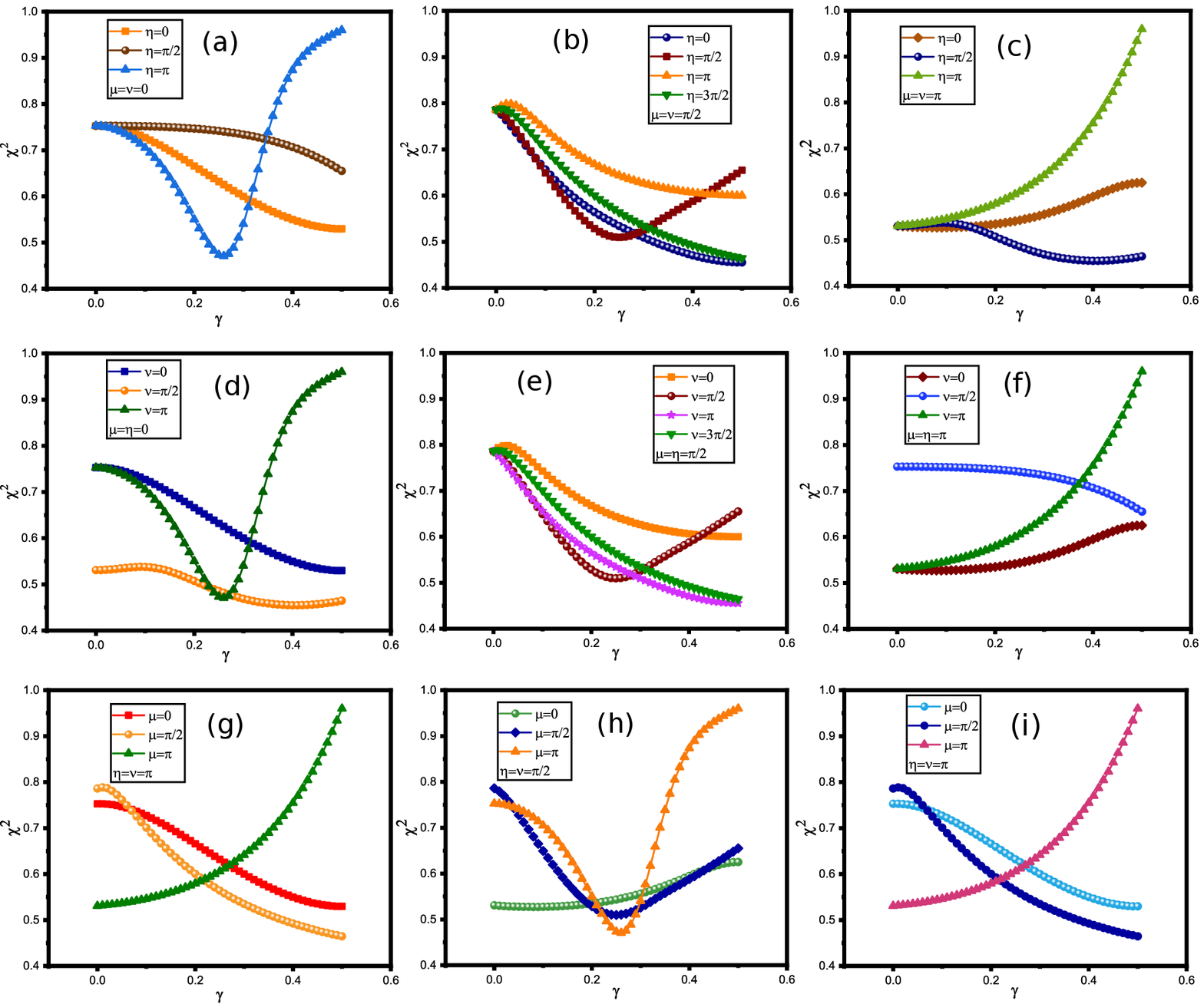}
	\caption{ Variation of $\chi^2$ with respect to $\gamma$, keeping $\alpha=\beta=\frac{1}{2}$ and for different phase values of $\mu,\nu,\eta$.}
	\label{phase_2}
\end{figure*}

We observed the variation of RMQFI with respect to phase parameters.  Initially we considered a pure state keeping $\alpha$ $\beta$ at $\frac{1}{2}$ and varied other parameters $\delta$ and $\gamma$. We extended our study of RMQFI for phase parameters along with $\delta$, $\gamma$ too. We kept two phases fixed at different values 0, $\pi/2$ and $\pi$ and varied another phase along with the other two parameters. Fig -\ref{phase_1}, and \ref{phase_2} exhibit the repetitive pattern of some plots with different phases. Phases $\mu$, $\nu$, $\eta$ and phase differences of these terms are responsible for this behaviour. 

It is observed that for any range and combination we always have the $\chi^2<1$ implying the entanglement region and having a maximum value of RMQFI as 0.95 for a particular combination of equal probable state(except for the cases of (b) and (e) of figure \ref{phase_1} and \ref{phase_2}). {Different phase values of $\mu$, $\nu$, and $\eta$; $\chi^2$ will be different; that can be experimentally beneficial to obtain a wide range of $\chi^2$}. \\

\section{Conclusions}
In conclusion, we have studied QFI for general 3 qubit pure states in spin based system. We have observed different values of RMQFI, varying different parameters of the general states. Here we pointed out both graph and hypergraph lie in that entangled region. 
It is already discussed in this paper \cite{pezze2009entanglement} that the interferometer using classical separable states provides a very low phase estimation sensitivity than the special class of entangled states having $\chi<1$. The 3 qubit graph and hypergraph states discussed above are also eligible for this criteria: $F_Q>N$ which significantly says that they satisfy the necessary condition for the violation of Bell’s inequality \cite{frowis2019does} and allow sub shot noise (SSN) phase uncertainty \cite{krischek2011useful}. \\ We have prepared a general 3 qubit spin based quantum state \ref{gen}.   The lowest value of $\chi^2$ from the first case represents the graph state whereas the hypergraph state lies between the maximum and minimum of $\chi^2$. The two different structures of these states are responsible for different RMQFI. It is already discussed the importance of the graph states \cite{shettell2020graph} in the context of quantum metrology and we are also adding the hypergraph states here.\\

In the observation of graph and hypergraph states, we get the same values of $\chi^2$ while varying with different parameters. As the Graph and hypergraph states are equally probable with respect to those sub parts, this equal probability ensures the particular point in figure \ref{all}. We also found varying with different phases, some figures depict exactly the same patterns due to the phase difference of the configuration.\\

Oszmaniec et al. showed that most entangled states are
not good for quantum metrology, rather most symmetric states are useful for quantum metrology. Thus both graph and hypergraph states can be beneficial in this domain \cite{PhysRevX.6.041044}. {This 3- qubit graph state is one of the states which is maximally entangled and most symmetric.} 

Statistical speed for the graph state $\ket{G}$ and hypergraph state $\ket{H}$ are found to be $3$ and $2.165$ respectively. Other than GHZ state, this graph state $\ket{G}$ also reached the Heisenberg Scaling, $(\Delta\theta)^2|_{\theta=0}=1/N^2$ \cite{toth2014quantum}. {Attaining  Heisenberg limit(1/N) covers the broad area of applications from fundamental signals detection to experimental nonclassical states verification \cite{kura2020standard}.}\\

\end{document}